\documentclass[conference]{IEEEtran}
\IEEEoverridecommandlockouts
\usepackage{xcolor}
\usepackage{cite}
\usepackage{graphicx}
\usepackage{caption} 
\usepackage{subcaption} 
\usepackage[english]{babel} 
\usepackage{amsmath,amsfonts,amsthm,bm} 
\usepackage{wrapfig}
\usepackage{multirow}
\usepackage[framed,numbered,autolinebreaks,useliterate]{mcode}
\usepackage{mwe}
\usepackage{comment}
\usepackage{url} 
\usepackage{cite}

\ifCLASSINFOpdf
\else
\fi
\begin{document}
\title{A Compact Delay Model for OTS Devices}

\author{\IEEEauthorblockN{M. M. Al Chawa and R. Tetzlaff}
\IEEEauthorblockA{ 
 Technische Universität Dresden\\
Dresden, Germany\\
Email: mohamad\_moner.al\_chawa@tu-dresden.de}
\and
\IEEEauthorblockN{D. Bedau, J. W. Reiner, D. A. Stewart and M. K. Grobis}
\IEEEauthorblockA{ Western Digital San Jose Research Center\\San Jose, CA, USA\\
Email: daniel.bedau@wdc.com}
}

\maketitle

\begin{abstract}
This paper presents a novel compact delay model of Ovonic Threshold Switch (OTS) devices that works efficiently for circuit simulations. The internal state variable of the two terminal devices is estimated using a delay system  that uses a few electrical components related to a suggested equivalent circuit of the device. Finally, we tested the proposed model against measured data from devices fabricated by Western Digital Research. 
\end{abstract}

\section{Introduction}
Chalcogenide alloys offer the possibility to fabricate a family of two terminal Ovonic Threshold Switch (OTS) devices.
Usually, this kind of device presents a characteristic current-controlled negative differential resistance (NDR) on their DC $i\text{--}v$ curve. Due to this NDR, they exhibit an extremely fast and sharp switching between their on and off states, thus allowing them to be a good option for various applications: selectors for memory cells, fast switches, or neuromorphic computing\cite{tuma2016stochastic, song2019ovonic, readEvaluatingOvonicThreshold2021}.  

\begin{figure}[b!]
    \includegraphics[width=1.0\columnwidth, height=4cm]{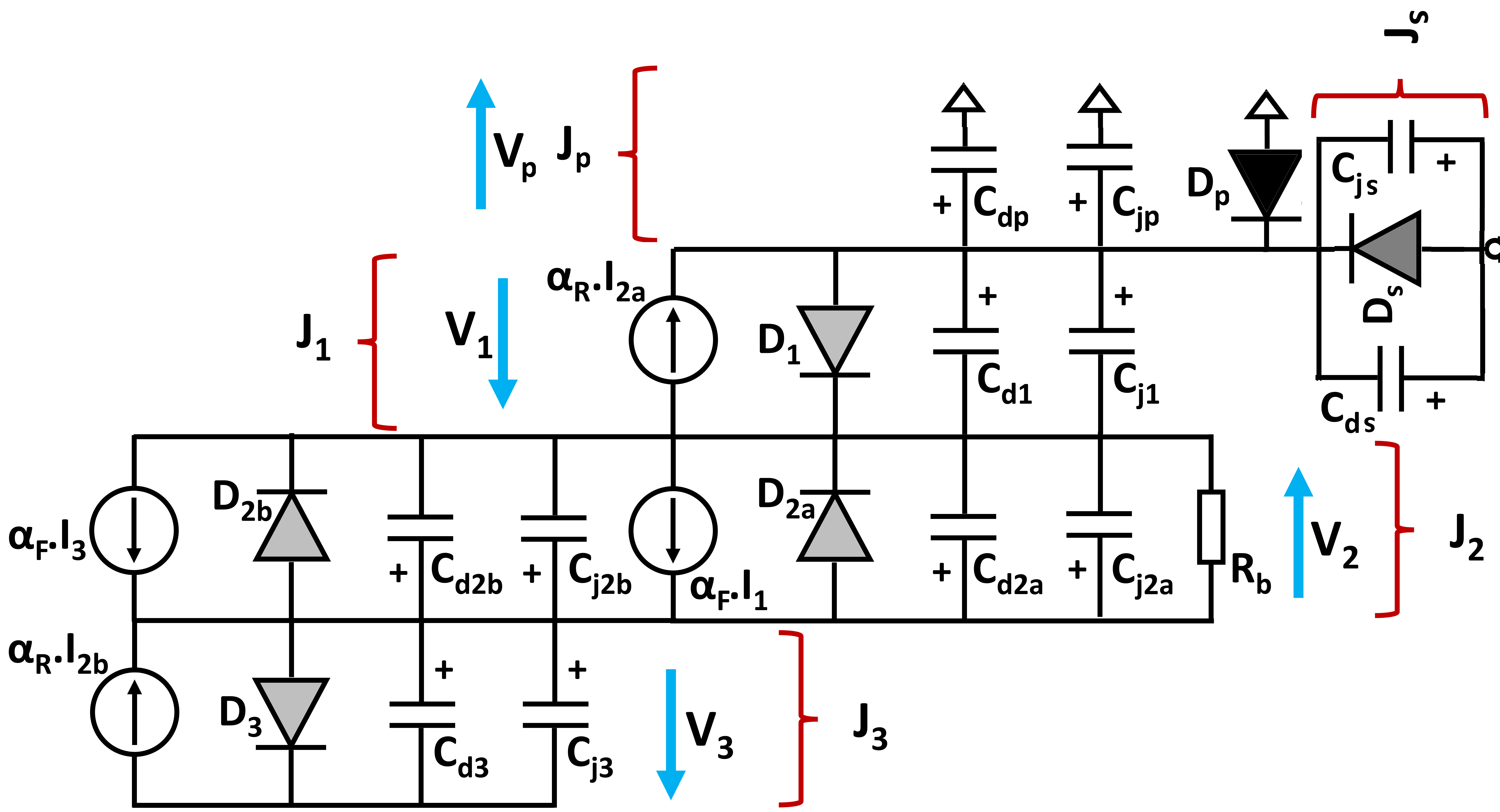}
    \caption{OTS cell equivalent circuit.  
    }
    \label{fig:newcircuit}
\end{figure}
 \begin{figure*}
     \centering
      \begin{subfigure}[b]{0.49\textwidth}
         \centering
                          \includegraphics[width=\textwidth]{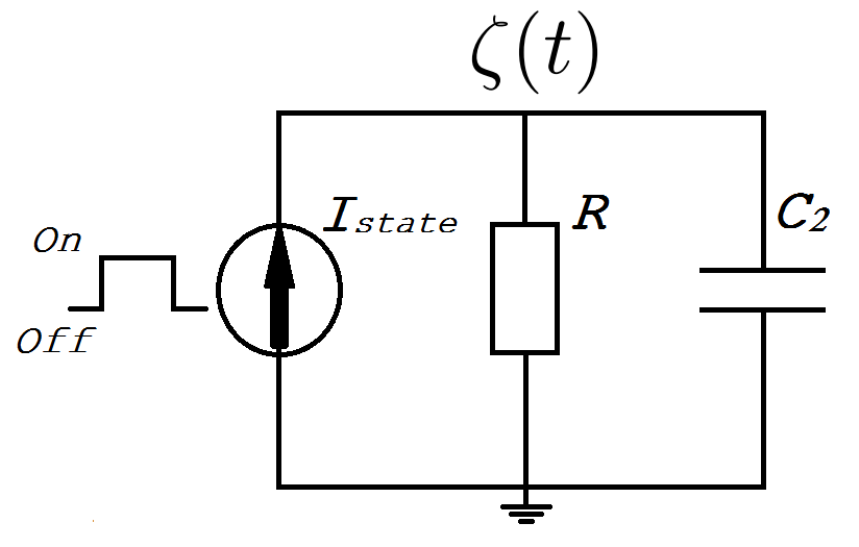}
         \caption{}
         \label{fig:a1}
     \end{subfigure}
     \hfill
     \begin{subfigure}[b]{0.5\textwidth}
         \centering
                    \includegraphics[width=\textwidth]{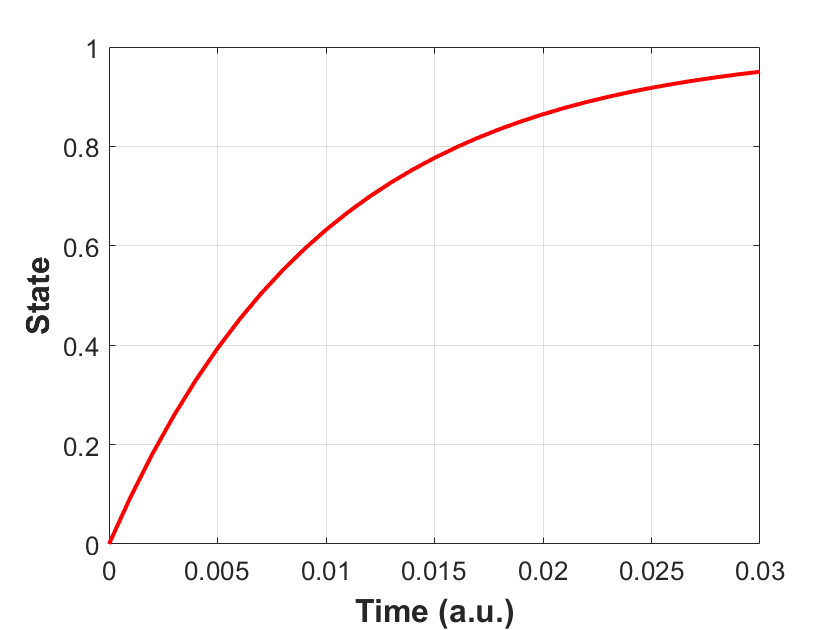}
         \caption{}
         \label{fig:state}
     \end{subfigure}
            \caption{
       The Delay model extracted from $J_2$ in Fig. \ref{fig:newcircuit} to obtain the internal state variable. (a) The proposed circuit related to the $J_2$ : $C_{2}=10nF$, $R_b=1M\Omega$ and $I_{state}=1\mu A$; (b) The values of the internal state variable (voltage drop across the circuit in Fig. \ref{fig:a1})
       vs. time (a.u.). }
        \label{fig:delay model}
\end{figure*}
\begin{figure*}
     \centering
      \begin{subfigure}[b]{0.496\textwidth}
         \centering
                  \includegraphics[width=\textwidth]{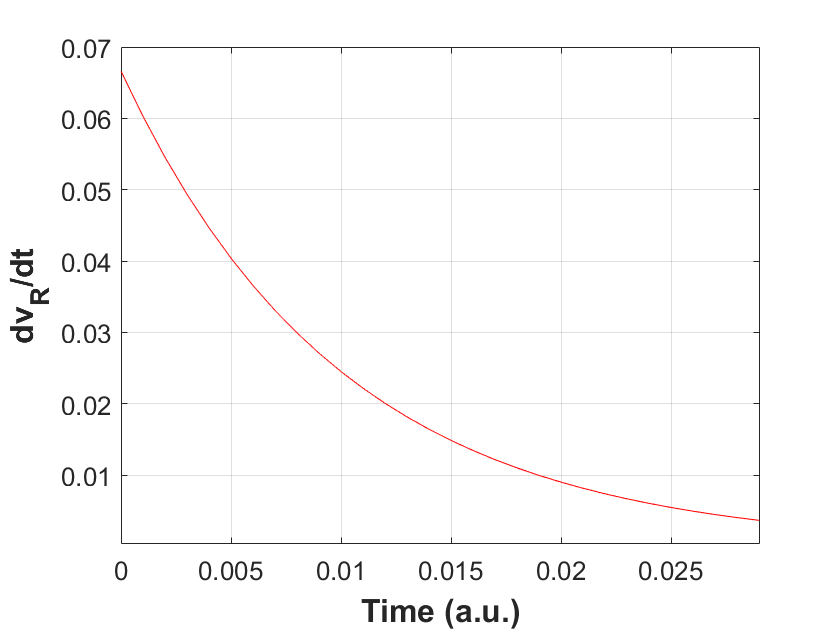}
         \caption{}
         \label{fig:dv}
     \end{subfigure}
     \hfill
     \begin{subfigure}[b]{0.496\textwidth}
         \centering
                    \includegraphics[width=\textwidth]{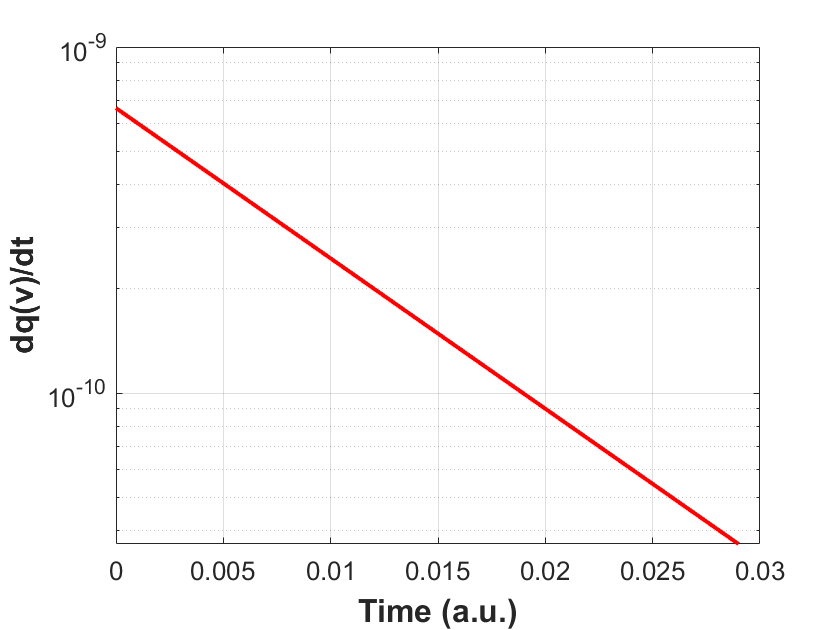}
         \caption{}
         \label{fig:charge model}
     \end{subfigure}
            \caption{
        (a) Derivative of the internal voltage drop $v_R$ across the $J_2$ 
        ($K=0.7$); (b) Capacitive current through $J_2$ ($v \approx v_R$) in \eqref{eq:QC} considering a linear capacitor approximation, ($ C \approx 10nF$).}
       
        \label{fig:Q}
\end{figure*}
\begin{figure*}
     \centering
      \begin{subfigure}[b]{0.496\textwidth}
         \centering
                  \includegraphics[width=\textwidth]{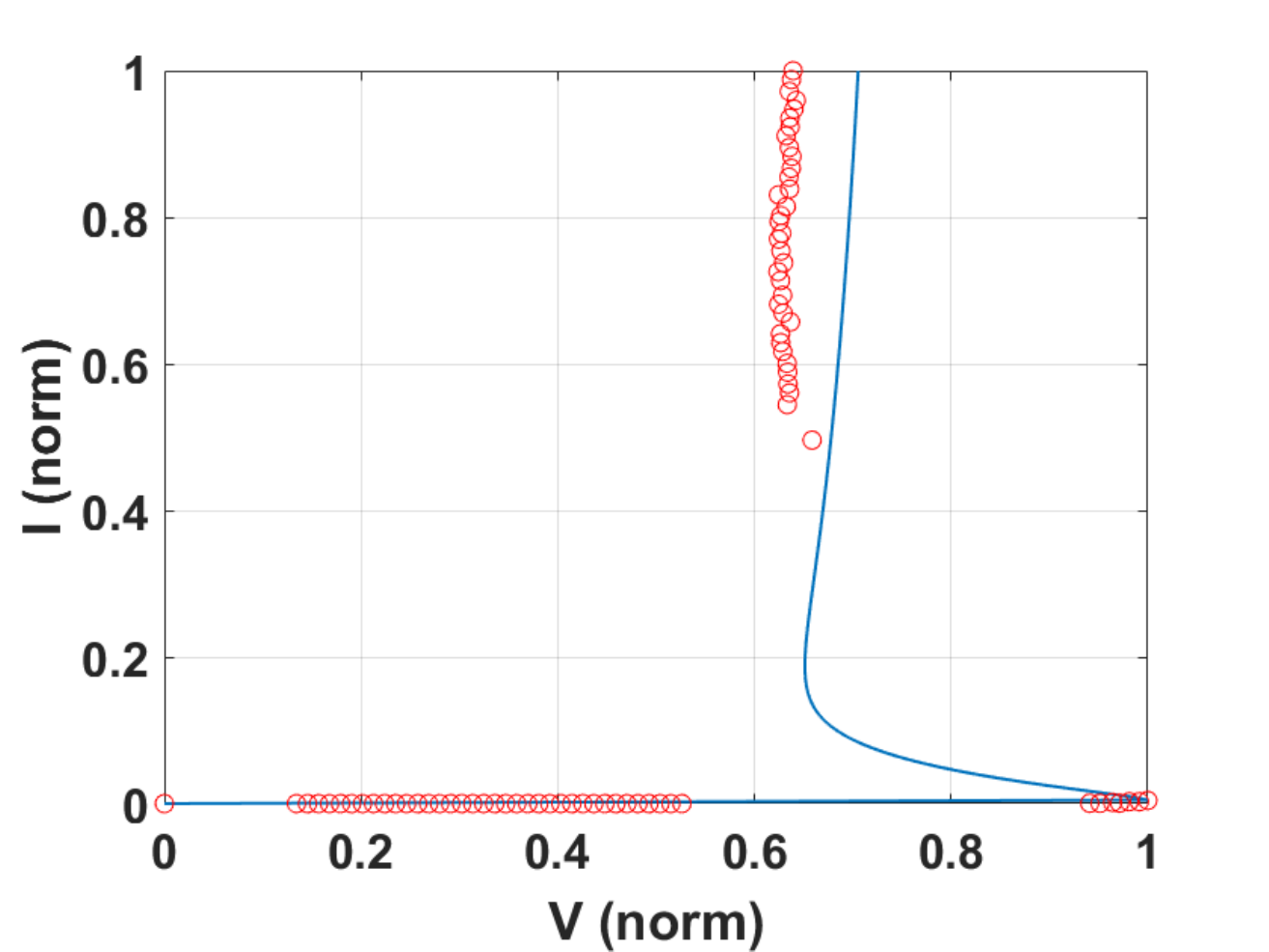}
         \caption{}
         \label{fig:Kless2}
     \end{subfigure}
     \hfill
     \begin{subfigure}[b]{0.496\textwidth}
         \centering
                    \includegraphics[width=\textwidth]{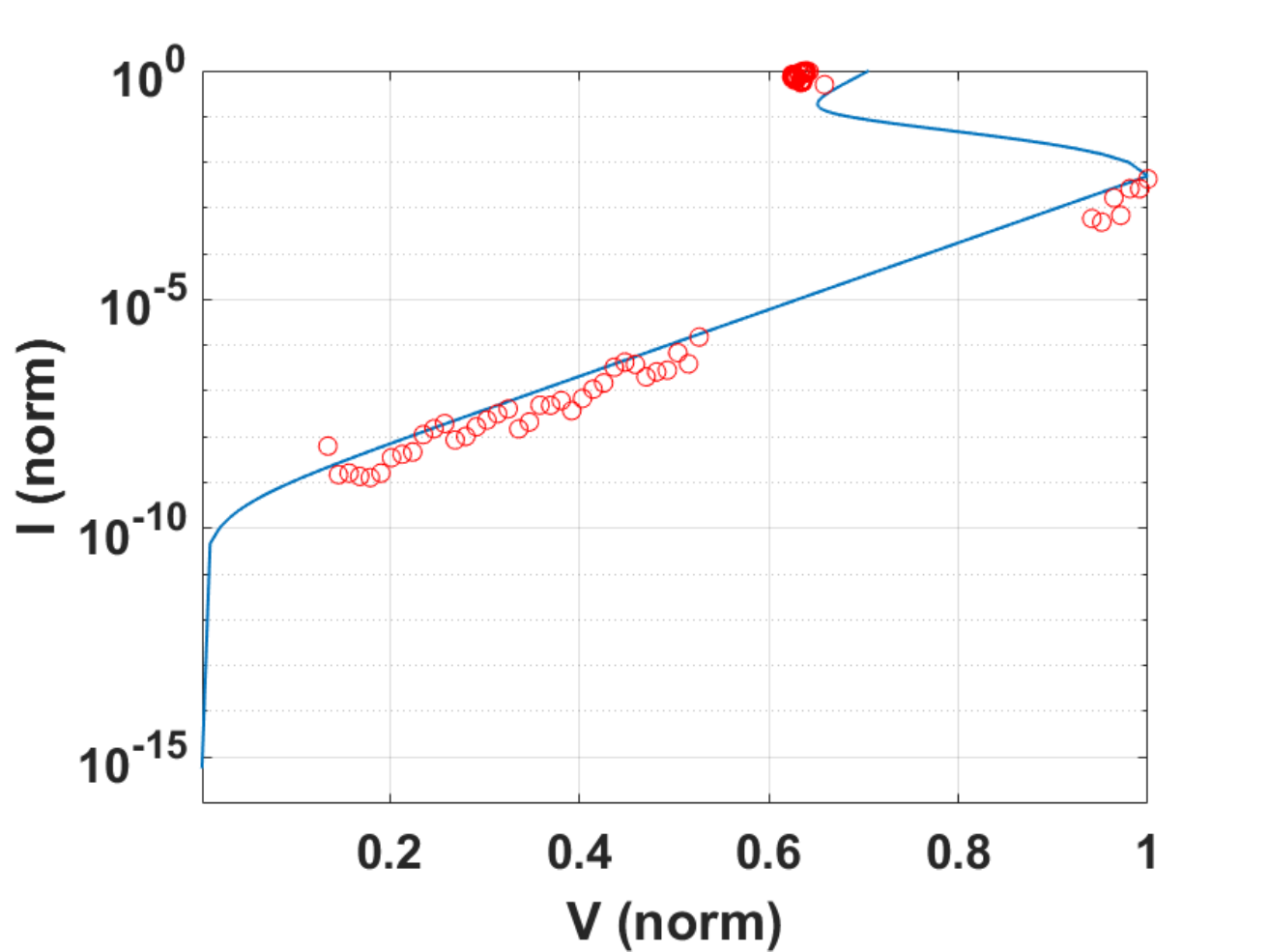}
         \caption{}
         \label{fig:Kmore2}
     \end{subfigure}
            \caption{
       $i\text{--}v$ curve, red dots are (normalized) experimental data (symbols) and modeled using \eqref{eq:final model} (blue line) for OTS devices. The devices and experimental measurements obtained by Western Digital Research\cite{10225437}.
       }
        \label{fig:fitting}
\end{figure*}

Even though the exact mechanism behind the threshold switching is still under discussion, many models can accurately reproduce the phenomenological conduct of OTS devices \cite{zhu_mrs_2019}, but they often lack efficiency when implemented on a circuit simulator. This is caused by the fact that these tools require a fast and precise model of the device described in a manner that suits an 
analogue simulation pipeline, including continuous derivatives. There is thus a compromise between the complexity that arises from the actual physical device and the required model accuracy. The capability of circuit simulation suites to run complicated models represents a challenge, that has been usually handled by the memristor device modelling community by using the classical current-voltage description \cite{9642394, 9629238}, or Chua's original proposition in flux and charge \cite{8357574,8713508, 9478939, 9181155, al2018exploring}. Any of these approaches are equivalent \cite{al2018exploring}, but following the most usual convention, we will use current and voltage in this work. 

Usual methods model the charge carrier dynamics utilising transport models\cite{9405114,9366291} that are described by differential equations. These physics-based models tend to be complicated, cannot be directly implemented in circuit simulators, and do not sufficiently serve as compact models for circuit simulations. 
An illustration of a non-physical model of a nano two-terminal device, ReRAM, for low computational cost as an alternative model, has been demonstrated in \cite{10.1145/3611315.3633237}.
However, in this work, the proposed model can be implemented in SPICE. Furthermore, a mathematical description of the equivalent circuit was obtained. The internal state variable, which controls the behaviour of the device, was extracted to get a compact model for OTS devices.  Finally, this model was tested to fit $i\text{--}v$ experimental data of a physical OTS device fabricated by Western Digital Research.

\section{Model Description and Implementation}
\label{sec:model}
\subsection{OTS Equivalent Circuit}
We have built an equivalent circuit for the OTS device shown in Fig. \ref{fig:newcircuit}. The circuit includes five junctions. A serial junction, $J_s$, and a parallel junction, $J_p$. Each junction consists of a diode and a capacitance in parallel.  
The other three junctions\cite{10225437},  $J_1$, $J_2$, and $J_3$.  Individually, the junction consists of a diode, a capacitance, and a current source in parallel. The diode is equal to the DC characteristic of the p-n junction current according to 
\begin{equation} 
  I_{J} =   \frac{I_s}{\alpha}\cdot (e^{v_J/V_{T}} -1)
  \label{J}
  \end{equation} 
  where,
  \begin{equation}
        \alpha =\frac{\beta}{1+ \beta}
        \label{eq:albe}
  \end{equation}
  $V_{\text{T}}$ is the thermal voltage $kT/q$ (
  $26 mV$  at $300 K$), and $I_{\text{S}}$ is the reverse saturation current.
  The current source is described as
\begin{equation} 
\alpha \cdot  I_{J}
\label{collect}
  \end{equation} 
The capacitance $ C_{total}= C_j + C_d$    \label{capacitane}
involves a junction capacitance
and a diffusion capacitance for each P-N junction (shown in Fig. \ref{fig:newcircuit}), and both capacitances are voltage-dependent.
We treat these junctions as Schottky junctions with the capacitance determined as\cite{6333317}
\begin{equation}
    C_{j}=\frac{C_{j0}}{(1-v_c/v_{j})^M},
\end{equation}
where $C_{j0}$ is the zero-bias capacitance, $v_{j}$ is the built-in potential, and $M$ corresponds to an experimental  coefficient.
The diffusion capacitance represents the minority carrier charge and is taken into account as
\begin{equation}
    C_{d} \approx \Sigma \tau_{j} \frac{I_j}{V_T}
\end{equation}
It is obvious that the proposed OTS equivalent circuit can be implemented and run in a circuit simulator.  
\subsection{Internal State Variable}
The behaviour of the two terminals OTS device is a function of the internal state variable. 
This state variable has been modelled as a delay function which goes from 0 to 1, and represents the switching of the state between its off and on values, respectively, as shown in Fig. \ref{fig:a1}. The delay model has been constructed as the circuit shown in Fig. \ref{fig:newcircuit} employing a few elements linked to the second junction. The capacitor $C_2$ is common for all the capacitors present in parallel in the second junction $J_2$.  The resistor $R_2$ is related to the diodes and internal bias resistor, while the current source $I_{State}$ is connected to the current sources in Fig. \ref{fig:newcircuit}. 
The current source in Fig \ref{fig:a1} is defined as two different values: it is assigned to $1 \mu A$ when the voltage across the device overreaches or catches the threshold voltage, $v_{th}$ (where $i_{th}$ is the corresponding current). Otherwise, the value of the current source is zero when the voltage across the device falls under the threshold voltage. Applying KCL, the delay model can be described as 
\begin{equation}
    I_{State}= \frac{v_{R}}{R_2} + C_2 \cdot \frac{dv_R}{dt} 
\end{equation}
Let us suppose an internal state variable (in Volt), $State$ or $  \zeta  \in [0,1]$, which can be obtained by solving the last equation as follows
\begin{equation}
    \zeta(t) =  I_{State} \cdot R_2 - e^{\frac{-t}{R_{2}C_2}} 
    \label{State}
\end{equation}
 The plot of \eqref{State} is shown in Fig. \ref{fig:state}.   
 Assuming a scaling factor $K$, the internal voltage drop across the $J_2$ can be expressed as   
\begin{equation}
  v_R=  K \cdot  \zeta 
  \label{KState}
\end{equation}
 Fig. \ref{fig:dv} illustrates the derivative of the last equation which represents the derivative of the internal voltage drop across the bias resistor.
%
%
 \subsection{Model Implementation}
 The proposed equivalent circuit in  Fig. \ref{fig:newcircuit} can be implemented and run in a circuit simulator, which expects the behaviour by solving a mathematical model of the equivalent circuit. The model involves the compact models of the elements and their topological connectivity that form the circuit, including the capacitance. The dynamic of the mathematical model of the circuit is provided by\cite{7154394, 10225437}
 \begin{equation}
  i(v)= \frac{dq(v)}{dt} + f(v)
     \label{eq:i12}
      \end{equation}
where $f$ is the currents passing through static branches of the equivalent circuit, and $dq/dt$ is the (capacitive) current passing through time-dependent branches of the equivalent circuit. 
We already have defined the voltage drop across the second junction, $J_2$, as the internal state variable of the device, $v_2=-v_R$. The current passing through the time-dependent branch connected to $v_2$ and $J_2$ is the main branch and can be given as
\begin{multline}    
\frac{dq(v)}{dt} \approx  \frac{dq(v_2)}{dt} = \frac{d [C_{total2} \cdot v_2]}{dt}
\approx -~C \cdot \frac{dv_R}{dt} 
 \label{eq:QC}
\end{multline}
As an approximation, a linear capacitor has been considered $C_{total2}$ $\approx$ $C$, as we focus particularly on the modelling of $i\text{--}v$ snapback in this contribution. 
A better proper charge model described in \eqref{capacitane} can be employed in this formulation without changing the equation structure. Fig. \ref{fig:charge model} illustrates the  current passing through time-dependent branches of equivalent capacitance, $10nF$ in \eqref{eq:QC}.
Also, It should be known, that the voltage drop across the series junction $J_s$ is negligible, $v_s=0$ . As a result of this, the voltage drop across the parallel junction $J_p$ is approximated to $v_p=-v$, where $v$ is the applied voltage across the OTS device. The first junction $J_1$ and the third junction  $J_3$ are identical. So, the voltage drop across the first junction, $J_1$, is equal to the voltage drop across the third junction, $J_3$, i.e. $v_1=v_3$. As a result of this, the voltage drop across the first junction, $J_1$, or the third junction, $J_3$, can be expressed as follows  
\begin{equation}
    v_1=v_3=\frac{v+v_R}{2}
    \label{eq:v1v3}
\end{equation}
On the other hand, the currents passing through the static branches can be obtained by using KCL as follows
\begin{equation} 
  f(v)= I_1 - \alpha_{R} \cdot  I_2  - I_P
  \end{equation} 
replacing \eqref{J}, \eqref{eq:albe}, and \eqref{eq:v1v3} 
in the  last equation
yielding to  
\begin{multline}
     f(v)= I_s \cdot [e^{v+v_{R}/2 \cdot V_T} (1 + \frac{1}{\beta_F}) -e^{-v_{R}/V_T} - \frac{1}{\beta_F}] - 
     \\ [\frac{I_s}{\alpha_R}\cdot (e^{-v/V_{T}} -1)]
\label{dc_final1}
\end{multline}
%
\begin{table}[]
\renewcommand{\arraystretch}{1.3}
\centering
\caption{The parameter values used to fit experimental measurements in Fig. \ref{fig:fitting} using \eqref{eq:final model}
.}
\label{tab:parameter}
\begin{tabular}{|c|c|}
\hline
\textbf{Parameter} & \textbf{Value} \\ \hline
      $I_s$    &     $10^{-14} A$      \\ \hline
      $\beta_F$    &   $250$        \\ \hline
        $\alpha_R$    &   $1$        \\ \hline
      $V_T$    &     $25.9mV$      \\ \hline
       $K$   &   $0.7$        \\ \hline
       $I_{State}$   & $1 \mu A$          \\ \hline
      $R$    &    $1M \Omega$       \\ \hline
       $C$   &     $10nF$      \\ \hline
       $v_{th}$   &   $2.4V$        \\ \hline
       $i_{th}$   &    $1 \mu A$        \\ \hline
         $R_{b}$   &     $5 k \Omega$         \\ \hline
\end{tabular}
\end{table}
Eventually, the current $i$ of the OTS device in \eqref{eq:i12} can be integrated into one formula using 
\eqref{eq:QC}, and \eqref{dc_final1} as follows
\begin{multline}
      i(v, v_R) = I_s \cdot [e^{v+v_{R}/2 \cdot v_T} (1 + \frac{1}{\beta_F}) -e^{-v_{R}/v_T} - \frac{1}{\beta_F} - 
     \\ \frac{1}{\alpha_R}\cdot (e^{-v/V_{T}} -1)]  -C \cdot \frac {d}{dt}v_R  
  \label{eq:final model}
\end{multline}
  We have tested the model with experimental data for a physical OTS device, which consisted of a Se-based OTS layer, approximately 15nm thick, with carbon electrodes. The OTS layer and electrodes were patterned into a pillar of approximately 40nm diameter.  Data was obtained by using a voltage pulse, with an on-chip resistor used to limit the current after OTS thresholding.
The parameter values for this fitting are listed in Table \ref{tab:parameter}. By overlaying the fitting line with the measurement, we show that the model reproduces the $i\text{--}v$ characteristics nicely, not just in the on and off states, but also in the snapback region as can be noticed in Fig. \ref{fig:fitting}. 
      \section{Conclusion}
\label{sec:discussion}
This work presents a novel compact delay model for OTS devices.  The internal state variable of the two terminal devices is evaluated utilising a delay circuit with few electrical components related to the proposed equivalent circuit of the OTS cell. 
Lastly, we validated the  model by fitting $i\text{--}v$ experimental data of the OTS devices fabricated by Western Digital Research.  

\end{document}